\documentstyle[prl,aps,psfig,epsf,twocolumn]{revtex}
\begin{document}
\draft
\twocolumn[\hsize\textwidth\columnwidth\hsize\csname@twocolumnfalse%
\endcsname

\preprint{}

\title{Density Wave States of Non-Zero Angular Momentum}
\author{Chetan Nayak}
\address{Physics Department, University of California, Los Angeles, CA 
  90095--1547}
\date{\today}
\maketitle

\begin{abstract}
We study the properties of states in which
particle-hole pairs of non-zero angular momentum
condense. These states generalize charge- and
spin-density-wave states, in which $s$-wave
particle-hole pairs condense.
We show that the $p$-wave spin-singlet
state of this type has Peierls ordering, while
the $d$-wave spin-singlet state is the staggered
flux state. We discuss model Hamiltonians which
favor $p$-wave and $d$-wave density wave order.
There are analogous orderings for pure spin models,
which generalize spin-Peierls order.
The spin-triplet density wave states
are accompanied by spin-$1$ Goldstone bosons,
but these excitations do not contribute to the
spin-spin correlation function. Hence, they must be
detected with NQR or Raman scattering experiments.
Depending on the geometry and topology of the Fermi surface,
these states may admit gapless fermionic
excitations. As the Fermi surface geometry
is changed, these excitations disappear
at a transition which is third-order in
mean-field theory.
The singlet $d$-wave and triplet $p$-wave density
wave states are separated from the
corresponding superconducting states
by zero-temperature $O(4)$-symmetric critical points.
\end{abstract}
\vspace{1 cm}

\vskip -0.4 truein
\pacs{PACS numbers:71.10.Hf, 75.10.Lp, 75.30.Fv, 71.27.+a,}
]
\narrowtext
%\newpage

\section{Introduction}

In recent years, a number of materials have been
uncovered in which the competition between
an effective attractive interaction and short-range
repulsion appears to lead to the formation of superconducting
states in which the Cooper pairs have non-zero relative angular
momentum. In this paper, we suggest that such competition
can also lead to density-wave states formed by the
condensation of {\it particle-hole}
pairs of non-zero relative angular momentum.
These states generalize the familiar charge- and
spin-density-wave states, in which $s$-wave
particle-hole pairs condense.
We discuss several different possible ordering schemes,
the types of interactions which favor them,
their physical properties, and their possible relevance
to experiments.

Several such states are already commonly known
by other names, as we will show below. The
singlet $l=1$ density-wave state is simply
the Peierls state (or bond-ordered wave), while the singlet $l=2$
density-wave state is known as the staggered flux state
of \cite{Kotliar88a,Marston89,Hsu91}.
However, the triplet analogues of these states have not been
discussed. Since the triplet analogues of these
states break spin-rotational invariance,
they have $S=1$ Goldstone boson excitations.
However, the ground state does not have a
non-zero expectation value for the spin at
any wavevector. Hence, as we will see,
these Goldstone bosons cannot be detected in
experiments which couple simply to the spin
density, such as neutron scattering or NMR.
Instead, Raman scattering or NQR are necessary
to couple to the Goldstone bosons of these more
subtle types of ordering. More generaly, $s$-wave
probes cannot couple directly to the orders
discussed here; instead, local probes or those
which couple to higher powers of the order parameter
are necessary.

The $p$-wave and $d$-wave density wave
states are favored by the same types of interactions
which favor the $s$-wave state -- i.e. the CDW.
However, they evade interactions which disfavor
CDW order. Similarly, they are favored by
superconductivity-favoring
pair-hopping terms \cite{Wheatley88,Chakravarty93,Hirsch89,Assaad97}
while evading interactions which disfavor superconductivity.
Hence, they are rather natural candidates for
systems with competing repulsive interactions.

As in the case of higher angular momentum
superconducting states, there is the possibility of
gapless excitations since the order parameter
can have nodes on the Fermi surface. To consider
one consequence of this, suppose that
the shape of the Fermi surface is such that
the nodes of the order parameter do not lie on the
Fermi surface. Let us distort the shape of the
Fermi surface by, say, changing the anisotropy
between the hopping parameters, which can be done
by applying uniaxial pressure. At the mean-field level,
a third-order phase transition can occur at which gapless excitations
appear. After this point, the system remains critical
as a result of the node.

The analogy with supercondutivity can be taken a step
further by combining density wave order with
superconducting order in a pseudospin $SU(2)$ triplet
following Yang \cite{Yang89} and Zhang \cite{Zhang97}.
At a critical point between the two types of order,
this pseudospin $SU(2)$ could become exact, giving -- together with $SU(2)$
spin symmetry -- an $O(4)$-invariant critical point.
We discuss the possible relevance of such a critical
point to the pseudogap regime of the cuprate superconductors.

Particle-hole condensates with non-zero
angular momentum were considered in 
the context of excitonic insulators
by Halperin and Rice \cite{Halperin68}.
They were rediscovered in the context of
the mean-field instabilities of of extended
Hubbard models by Schulz \cite{Schulz89b}
and Nersesyan \cite{Nersesyan91a,Nersesyan91b}
and collaborators. At around the same time,
Kotliar \cite{Kotliar88a} and Marston and
Affleck \cite{Marston89} found the staggered flux state
as a mean-field solution of the Hubbard model.
However, it was apparently not recognized that
the singlet $d_{{x^2}-{y^2}}$ density-wave state
is the same as the staggered flux state. More recently,
this state \cite{Nayak95,Nayak97,Chakravarty00} and a related
variant \cite{Varma97,Varma99} have been discussed in the context
of the cuprate superconductors. A version
of this state (see the comments in the concluding
section) has appeared in mean field analyses
of an $SU(2)$ mean field theory of the $t-J$
model \cite{Wen96,Lee98}. The Nodal Liquid state
of \cite{Balents98,Balents99a,Balents99b}
also bears a family resemblance to the staggered flux
state; we will return to the relationship between these
states in the concluding section.

\section{Order Parameters and Broken Symmetries}

We define the different possible density-wave
orderings by analogy with the more familiar
superconducting case.
Consider a system of electrons on a square
lattice of side $a$. A superconductor is defined
by a non-vanishing
expectation value of
\begin{equation}
\left\langle {\psi_\alpha}(k,t)\,
{\psi_\beta}(-k,t) \right\rangle
\end{equation}
A triplet superconductor is characterized by
the expectation value
\begin{equation}
\left\langle {\psi_\alpha}(k,t)\,
{\psi_\beta}(-k,t) \right\rangle
= \vec{\Delta}(p)\cdot
{\vec{\sigma}_\alpha^{\,\,\,\gamma}}
{\epsilon_{\gamma\beta}}
\end{equation}
Fermi statistics requires that $\vec{\Delta}(p)$
be odd in $\vec{p}$. $p$-wave superconductors can have
the components of $\vec{\Delta}(p)$ chosen from
$\sin {k_x}a$, $\sin {k_y}a$, or
$\sin {k_x}a \pm  i \sin {k_y}a$.
For instance, a $p_x$ superconductor
with all spins polarized along the $3$-direction
will have ${\Delta_1}+i{\Delta_2}\neq 0$
and ${\Delta_3} = {\Delta_1}-i{\Delta_2} = 0$:
\begin{equation}
\left\langle {\psi_\alpha}(k,t)\,
{\psi_\beta}(-k,t) \right\rangle
= {\Delta_0}\,\left(\sin {k_x}a\right) \,
{{\sigma}_\alpha^{+\,\,\,\gamma}}
{\epsilon_{\gamma\beta}}
\end{equation}
Spin-polarized $p_y$ and ${p_x}+i{p_y}$ superconductors
have $\sin {k_x}a$ replaced, repectively,
by $\sin {k_y}a$ and $\sin {k_x}a + i \sin {k_y}a$.
The analog of the $A'$ phase of $^3$He has
equal numbers of $\uparrow\uparrow$ and
$\downarrow\downarrow$ pairs:
\begin{eqnarray}
\left\langle {\psi_\alpha}(k,t)\,
{\psi_\beta}(-k,t) \right\rangle
&=& \cr & & {\hskip - 1 cm}
{\Delta_0}\,\left(\sin {k_x}a \,\,
{{\sigma}_\alpha^{1\,\,\,\gamma}}
+ \sin {k_y}a \,\, {{\sigma}_\alpha^{2\,\,\,\gamma}}
\right) {\epsilon_{\gamma\beta}}
\end{eqnarray}
An unpolarized $p_x$ superconductor of
$\uparrow\downarrow$ pairs has
${\Delta_3}\neq 0$ and ${\Delta_1}={\Delta_2} = 0$:
\begin{equation}
\left\langle {\psi_\alpha}(k,t)\,
{\psi_\beta}(-k,t) \right\rangle
= {\Delta_0}\,\left(\sin {k_x}a\right) \,
{{\sigma}_\alpha^{3\,\,\,\gamma}}
{\epsilon_{\gamma\beta}}
\end{equation}
As in the polarized case, unpolarized
$p_y$ and ${p_x}+i{p_y}$ superconductors
have $\sin {k_x}a$ replaced, repectively,
by $\sin {k_y}a$ and $\sin {k_x}a + i \sin {k_y}a$.
In principle, more complicated order parameters are
possible, with all components of $\vec{\Delta}$
taking non-vanishing values. If any component
of $\vec{\Delta}(p)$ is not real, time-reversal
symmetry ($T$) is broken.

A $d$-wave superconductor must
be a spin-singlet superconductor. A
$d_{{x^2}-{y^2}}$ superconductor has
\begin{equation}
\left\langle {\psi_\alpha}(k,t)\,
{\psi_\beta}(-k,t) \right\rangle
= {\Delta_0}\,\left( \cos {k_x}a - \cos {k_y}a\right)\,
{\epsilon_{\alpha\beta}}
\end{equation}
while a $d_{xy}$ superconductor has
$\cos {k_x}a - \cos {k_y}a$ replaced
by $\sin {k_x}a\,\sin {k_y}a$.
A ${d_{{x^2}-{y^2}}}+i{d_{xy}}$ superconductor
breaks $T$ with the order parameter:
\begin{eqnarray}
\left\langle {\psi_\alpha}(k,t)\,
{\psi_\beta}(-k,t) \right\rangle
= {\hskip 4.3 cm}\cr {\Delta_0}\left( \cos {k_x}a - \cos {k_y}a
+ i \sin {k_x}a\,\sin {k_y}a\right)
{\epsilon_{\alpha\beta}}
\end{eqnarray}

We can define analogous orders for density-wave states.
However, the spin structures will no longer
be determined by Fermi statistics.
Let us first consider the singlet orderings.
A singlet $s$-wave density wave is simply
a charge-density-wave\footnote{Extended $s$-wave
is also possible.}:
\begin{equation}
\left\langle {\psi^{\alpha\dagger}}(k+Q,t)\,
{\psi_\beta}(k,t) \right\rangle
= {\Phi_Q}\,\, {\delta^\alpha_\beta}
\end{equation}
A singlet $p_x$ density-wave state
has ordering
\begin{equation}
\left\langle {\psi^{\alpha\dagger}}(k+Q,t)\,
{\psi_\beta}(k,t) \right\rangle
= {\Phi_Q} \sin {k_x}a\,\,
{\delta^\alpha_\beta}
\end{equation}
The singlet ${p_x}+i{p_y}$ density-wave states
are defined by:
\begin{equation}
\left\langle {\psi^{\alpha\dagger}}(k+Q,t)\,
{\psi_\beta}(k,t) \right\rangle
= {\Phi_Q}\,
\left(\sin {k_x}a + i \sin {k_y}a\right)\,\,
{\delta^\alpha_\beta}
\end{equation}

Similarly, the singlet $d_{{x^2}-{y^2}}$
density-wave states have
\begin{equation}
\left\langle {\psi^{\alpha\dagger}}(k+Q,t)\,
{\psi_\beta}(k,t) \right\rangle =
{\Phi_Q}\left( \cos {k_x}a - \cos {k_y}a\right)\,\,
{\delta^\alpha_\beta}
\end{equation}
while the singlet ${d_{{x^2}-{y^2}}}+i{d_{xy}}$
density-wave states have
\begin{eqnarray}
\left\langle {\psi^{\alpha\dagger}}(k+Q,t)\,
{\psi_\beta}(k,t) \right\rangle = {\hskip 4.3 cm}\cr
{\Phi_Q}\,\left( \cos {k_x}a - \cos {k_y}a
+ i \sin {k_x}a\,\sin {k_y}a\right)\,\,
{\delta^\alpha_\beta}
\end{eqnarray}

These states belong to a class of states of the form:
\begin{equation}
\left\langle {\psi^{\alpha\dagger}}(k+Q,t)\,
{\psi_\beta}(k,t) \right\rangle =
{\Phi_Q}\,f(k)\,\,
{\delta^\alpha_\beta}
\end{equation}
$f(k)$ is an element of
some representation of the space group
of the vector $\vec{Q}$ in the square lattice.
In this paper, we will focus primarily
on the cases $f(k) =\sin {k_x}a $
and $f(k) = \cos {k_x}a - \cos {k_y}a$, but
$f(k)$ could be an element of
some larger representation. The $s$-wave
(or extended $s$-wave) cases, $f(k)=|f(k)|$,
are the usual charge-density wave states.

$Q$ is the wavevector at which the density-wave ordering
takes place. It may be commensurate or
incommensurate\footnote{In this paper,
we will take commensurate to mean the
situation in which $2\vec{Q}$ is a reciprocal lattice vector.
The term `incommensurate' will actually include higher-order
commensurability.}.
For commensurate ordering such that $2Q$ is a reciprocal
lattice vector, e.g. $Q=(\pi/a,0)$ or $Q=(\pi/a,\pi/a)$,
we can take the hermitian conjugate
of the order parameter:
\begin{eqnarray}
\label{eqn:comm_con}
\left\langle {\psi^{\dagger\beta}}(k,t)\,
{\psi_{\alpha}}(k+Q,t)
\right\rangle &=&
{\Phi_Q^*}\,f^*(k)\,{\delta^\alpha_\beta}\cr
\left\langle {\psi^{\beta\dagger}}(k+Q+Q,t)\,
{\psi_{\alpha}}(k+Q,t)
\right\rangle &=&
{\Phi_Q^*}\,f^*(k)\,{\delta^\alpha_\beta}\cr
{\Phi_Q}\,f(k+Q)\,{\delta^\alpha_\beta}
&=& {\Phi_Q^*}\,f^*(k)\,{\delta^\alpha_\beta}
\end{eqnarray}
Therefore, for $Q$ commensurate
\begin{equation}
\label{eqn:comm_phase}
\frac{f(k+Q)}{f^*(k)} = \frac{\Phi_Q^*}{\Phi_Q}
\end{equation}
Hence, if $f(k+Q)=-f^*(k)$, $\Phi_Q$ must be imaginary.
For singlet $p_x$ ordering, this will be the case if
$Q=(\pi/a,0)$ or $Q=(\pi/a,\pi/a)$. For
singlet $d_{{x^2}-{y^2}}$ ordering, this will be the case if
$Q=(\pi/a,\pi/a)$. If $f(k+Q)=f^*(k)$, $\Phi_Q$
must be real. For singlet $p_x$ ordering, this will be the case if
$Q=(0,\pi/a)$. For singlet $d_{xy}$ ordering, this will be
the case if $Q=(\pi/a,\pi/a)$.

For incommensurate ordering, $\Phi_Q$
can have arbitrary phase: the phase of 
$\Phi_Q$ is the Goldstone boson of broken translational
invariance, i.e. the sliding density-wave mode.
Impurities will pin this mode -- at second order in the
impurity potential, as in the
case of a spin-density-wave --
so we will not consider it further.

All of these states break translational
and rotational invariance. To further analyze the symmetries
of these states, it is instructive to write these orderings in
real space. The singlet $p_x$ density-waves have
non-vanishing expectation value:
\begin{eqnarray}
\left\langle {\psi^{\dagger\alpha}}(\vec{x},t)\,
{\psi_{\beta}}(\vec{x}+a{\hat x},t) - 
{\psi^{\dagger\alpha}}(\vec{x},t)\,
{\psi_{\beta}}(\vec{x}-a{\hat x},t)
\right\rangle = \cr
 \ldots -\frac{i}{2} \left({\Phi_Q}\,{e^{i\vec{Q}\cdot\vec{x}}}
+{\Phi_{-Q}}\,{e^{-i\vec{Q}\cdot\vec{x}}}\right)
{\delta^\alpha_\beta}
\end{eqnarray}
We have only written the modulated term;
the $\ldots$ refers to the uniform contribution
coming from the Fourier transform of
${\psi^\dagger}(k)\psi(k)$.

\begin{figure}[htb]
\centerline{\psfig{figure=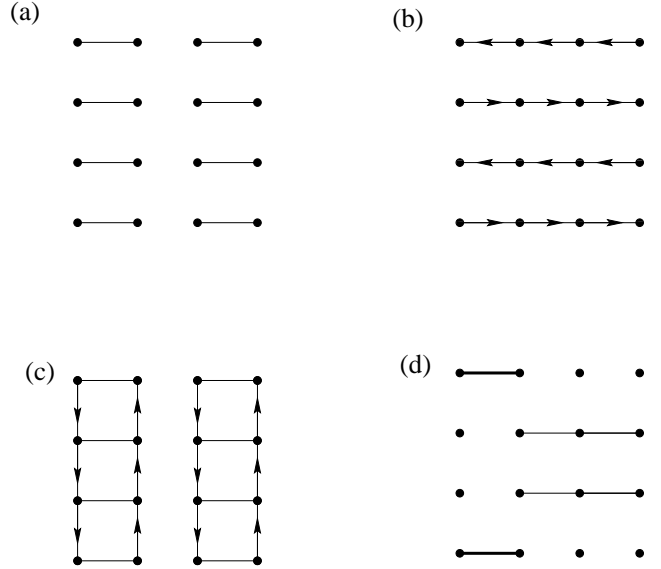,height=3in}}
\vskip 0.5cm
\caption{(a) $Q=(\pi/a,0)$ $p_x$ density-wave state.
(b) $Q=(0,\pi/a)$ $p_x$ density-wave state.
(c) $Q=(\pi/a,0)$ ${p_x}+i{p_y}$ density-wave state.
(d) Incommensurate $p_x$ density-wave state.
Arrowless lines are bonds where the kinetic energy
is large but there is no net current. Line thickness
indicates bond strength. Arrowed lines
denote currents.}
\label{fig:cpdw}
\end{figure}

Let us consider the commensurate and
incommensurate cases separately.
The incommensurate singlet $p_x$ density-wave states completely break
the translational and rotational symmetries.
If ${\Phi_Q}=-{\Phi_{-Q}^*}$,
$T$ is preserved; otherwise, it is broken.
The singlet ${p_x}+i{p_y}$ density-wave states 
always break $T$.
The commensurate states, on the other hand,
break translation by one lattice spacing;
translation by two lattice
spacings is preserved.
From (\ref{eqn:comm_phase}), a commensurate
singlet $p_x$ density-wave state
with $Q=(\pi/a,0)$ must have imaginary
$\Phi_Q$:
\begin{eqnarray}
\left\langle {\psi^{\dagger\alpha}}(\vec{x},t)\,
{\psi_{\beta}}(\vec{x}+a{\hat x},t) - 
{\psi^{\dagger\alpha}}(\vec{x},t)\,
{\psi_{\beta}}(\vec{x}-a{\hat x},t)
\right\rangle = \cr  \ldots + \left|{\Phi_Q}\right|\,
{e^{i\vec{Q}\cdot\vec{x}}}\,
{\delta^\alpha_\beta}
\end{eqnarray}
The singlet state of this type breaks
no other symmetries; it is usually called
the {\it Peierls state} or bond order wave.
If $Q=(0,\pi/a)$, $\Phi_Q$ must be real.
\begin{eqnarray}
\left\langle {\psi^{\dagger\alpha}}(\vec{x},t)\,
{\psi_{\beta}}(\vec{x}+a{\hat x},t) - 
{\psi^{\dagger\alpha}}(\vec{x},t)\,
{\psi_{\beta}}(\vec{x}-a{\hat x},t)
\right\rangle = \cr  \ldots - i \left|{\Phi_Q}\right|\,
{e^{i\vec{Q}\cdot\vec{x}}}\,
{\delta^\alpha_\beta}
\end{eqnarray}
As a result of the $i$, the $Q=(0,\pi/a)$
singlet $p_x$ density-wave states break $T$.
However, the combination of $T$ and translation
by an odd number of lattice spacings remains
unbroken. The same is true of the commensurate
singlet ${p_x}+i{p_y}$ density-wave states.
Examples of commensurate and incommensurate
singlet $p_x$ and  ${p_x}+i{p_y}$ density-wave states
are depicted in figure \ref{fig:cpdw}.

The singlet $d_{{x^2}-{y^2}}$ density-wave states have
non-vanishing expectation value:
\begin{eqnarray}
\left\langle {\psi^{\dagger\alpha}}(\vec{x},t)\,
{\psi_{\beta}}(\vec{x}+a{\hat x},t) + 
{\psi^{\dagger\alpha}}(\vec{x},t)\,
{\psi_{\beta}}(\vec{x}-a{\hat x},t)\right\rangle \cr
- \left\langle{\psi^{\dagger\alpha}}(\vec{x},t)\,
{\psi_{\beta}}(\vec{x}+a{\hat y},t) + 
{\psi^{\dagger\alpha}}(\vec{x},t)\,
{\psi_{\beta}}(\vec{x}-a{\hat y},t)
\right\rangle = \cr
 \ldots + \frac{1}{2}\left({\Phi_Q}\,{e^{i\vec{Q}\cdot\vec{x}}}
+{\Phi_{-Q}}\,{e^{-i\vec{Q}\cdot\vec{x}}}\right)
{\delta^\alpha_\beta}
\end{eqnarray}
The incommensurate singlet $d_{{x^2}-{y^2}}$ density-wave
states will preserve $T$ if ${\Phi_Q}={\Phi_{-Q}^*}$;
otherwise, they break $T$.
The same is true of the incommensurate singlet $d_{xy}$
density-wave states:
\begin{eqnarray}
\label{eqn:d_xy_def}
\left\langle {\psi^{\dagger\alpha}}(\vec{x},t)\,
{\psi_{\beta}}(\vec{x}+a{\hat x}+a{\hat y},t)\right\rangle + {\hskip 2 cm}\cr
\left\langle{\psi^{\dagger\alpha}}(\vec{x},t)\,
{\psi_{\beta}}(\vec{x}-a{\hat x}-a{\hat y},t)\right\rangle - {\hskip 2 cm}\cr
\left\langle {\psi^{\dagger\alpha}}(\vec{x},t)\,
{\psi_{\beta}}(\vec{x}-a{\hat x}+a{\hat y},t)\right\rangle - {\hskip 2 cm}\cr
\left\langle {\psi^{\dagger\alpha}}(\vec{x},t)\,
{\psi_{\beta}}(\vec{x}+a{\hat x}-a{\hat y},t)
\right\rangle \,\,\, = {\hskip 2 cm}\cr
 \ldots - \frac{1}{4}\left({\Phi_Q}\,{e^{i\vec{Q}\cdot\vec{x}}}
+{\Phi_{-Q}}\,{e^{-i\vec{Q}\cdot\vec{x}}}\right)
{\delta^\alpha_\beta}
\end{eqnarray}
Incommensurate singlet ${d_{{x^2}-{y^2}}} + i d_{xy}$
density-wave states necessarily break $T$:
\begin{eqnarray}
 \left\langle {\psi^{\dagger\alpha}}(\vec{x},t)\,
{\psi_{\beta}}(\vec{x}+a{\hat x},t) + 
{\psi^{\dagger\alpha}}(\vec{x},t)\,
{\psi_{\beta}}(\vec{x}-a{\hat x},t)\right\rangle \cr
-  \left\langle {\psi^{\dagger\alpha}}(\vec{x},t)\,
{\psi_{\beta}}(\vec{x}+a{\hat y},t) + 
{\psi^{\dagger\alpha}}(\vec{x},t)\,
{\psi_{\beta}}(\vec{x}-a{\hat y},t) \right\rangle \cr
+i \left\langle {\psi^{\dagger\alpha}}(\vec{x},t)\,
{\psi_{\beta}}(\vec{x}+a{\hat x}+a{\hat y},t)\right\rangle
{\hskip 3 cm} \cr
+i \left\langle{\psi^{\dagger\alpha}}(\vec{x},t)\,
{\psi_{\beta}}(\vec{x}-a{\hat x}-a{\hat y},t)\right\rangle
{\hskip 3 cm} \cr
-i \left\langle{\psi^{\dagger\alpha}}(\vec{x},t)\,
{\psi_{\beta}}(\vec{x}-a{\hat x}+a{\hat y},t)\right\rangle
{\hskip 3 cm} \cr
-i \left\langle{\psi^{\dagger\alpha}}(\vec{x},t)\,
{\psi_{\beta}}(\vec{x}+a{\hat x}-a{\hat y},t)
\right\rangle = {\hskip 2.5 cm} \cr
 \ldots + \left(\frac{1}{2}-\frac{i}{4}\right)
\left({\Phi_Q}\,{e^{i\vec{Q}\cdot\vec{x}}}
+{\Phi_{-Q}}\,{e^{-i\vec{Q}\cdot\vec{x}}}\right)
{\delta^\alpha_\beta}
\end{eqnarray}

\begin{figure}[htb]
\centerline{\psfig{figure=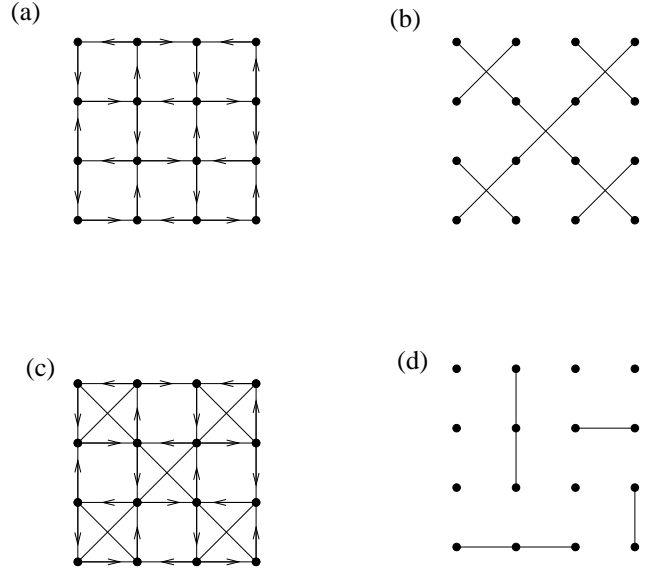,height=3in}}
\vskip 0.5cm
\caption{(a) Commensurate $d_{{x^2}-{y^2}}$ density-wave state.
(b) Commensurate $d_{xy}$ density-wave state.
(c) Commensurate ${d_{{x^2}-{y^2}}} + i d_{xy}$ density-wave state.
(d) Incommensurate, $T$-preserving $d_{{x^2}-{y^2}}$.
Arrowless lines are bonds where the kinetic energy
is large but there is no net current. Line thickness
indicates bond strength. Arrowed lines
denote currents.}
\label{fig:cddw}
\end{figure}

The commensurate $Q=(\pi/a,\pi/a)$
singlet $d_{{x^2}-{y^2}}$ density-wave states
must have imaginary ${\Phi_Q}$:
\begin{eqnarray}
\left\langle {\psi^{\dagger\alpha}}(\vec{x},t)\,
{\psi_{\beta}}(\vec{x}+a{\hat x},t)\right\rangle + 
\left\langle{\psi^{\dagger\alpha}}(\vec{x},t)\,
{\psi_{\beta}}(\vec{x}-a{\hat x},t)\right\rangle
- \cr \left\langle{\psi^{\dagger\alpha}}(\vec{x},t)\,
{\psi_{\beta}}(\vec{x}+a{\hat y},t)\right\rangle + 
\left\langle{\psi^{\dagger\alpha}}(\vec{x},t)\,
{\psi_{\beta}}(\vec{x}-a{\hat y},t)
\right\rangle = \cr  \ldots + \frac{i}{2}\left|{\Phi_Q}\right|\,
{e^{i\vec{Q}\cdot\vec{x}}}\,
{\delta^\alpha_\beta}
\end{eqnarray}
As a result of the $i$, the singlet $d_{{x^2}-{y^2}}$ density-wave
breaks $T$ as well as translational
and rotational invariance. The combination of time-reversal
and a translation by one lattice spacing is preserved
by this ordering.
The commensurate $Q=(\pi/a,\pi/a)$ singlet
$d_{{x^2}-{y^2}}$ density-wave
state is often called the {\it staggered flux state}.
There is also a contribution to this correlation function
coming from ${\psi^\dagger}(k)\psi(k)$ which is uniform
in space (the $\ldots$); as a result, the
phase of the above bond correlation
function -- and, therefore, the flux through each
plaquette -- is alternating.
The commensurate $Q=(\pi/a,\pi/a)$ singlet $d_{xy}$
must have real ${\Phi_Q}$; therefore,
it does not break $T$. On the other hand,
the singlet ${d_{{x^2}-{y^2}}} + i d_{xy}$ state does
break $T$.
Note that the nodeless commensurate
singlet ${d_{{x^2}-{y^2}}} + i d_{xy}$
density-wave state does not break more symmetries than the
commensurate singlet ${d_{{x^2}-{y^2}}}$ density-wave state, in
contrast to the superconducting case.
Examples of singlet $d_{{x^2}-{y^2}}$,
$d_{xy}$, and ${d_{{x^2}-{y^2}}} + i d_{xy}$
density-wave states are depicted in figure \ref{fig:cddw}.

We now consider the triplet density-wave states.
Triplet states all break spin-rotational
invariance and, therefore, have Goldstone boson
excitations. We will discuss the experimental
consequences of these Goldstone bosons later.
The triplet $s$-wave density wave state is simply
a spin-density-wave. The triplet $p$-wave
and $d$-wave states are characterized by:
\begin{equation}
\left\langle {\psi^{\alpha\dagger}}(k+Q,t)\,
{\psi_\beta}(k,t) \right\rangle
= {\vec{\Phi}_Q}(k) \cdot
{\vec{\sigma}^\alpha_\beta}
\end{equation}
with the components of ${\vec{\Phi}_Q}(k)$
chosen from, respectively, $\sin {k_x}a$, $\sin {k_y}a$,
$\sin {k_x}a \pm  i \sin {k_y}a$;
and $\cos {k_x}a - \cos {k_y}a$,
$\sin {k_x}a\,\sin {k_y}a$,
$\cos {k_x}a - \cos {k_y}a \pm i \sin {k_x}a\,\sin {k_y}a$.

A state in which the particle-hole pairs are polarized, which
is the most direct analogue of a spin-density-wave
has ${\vec{\Phi}_Q}(p)$ of the form ${\Phi_Q^3}\neq 0$,
${\Phi_Q^1}={\Phi_Q^2}= 0$:
\begin{equation}
\left\langle {\psi^{\alpha\dagger}}(k+Q,t)\,
{\psi_\beta}(k,t) \right\rangle
= {\Phi_Q}\,f(k)\,
{\sigma^{3\,\,\alpha}_\beta}
\end{equation}

\noindent
where $f(k)$ is chosen from the above set.
Alternatively, the particle-hole pairs can be unpolarized,
e.g.
\begin{equation}
\label{eqn:p-h_un_order}
\left\langle {\psi^{\alpha\dagger}}(k+Q,t)\,
{\psi_\beta}(k,t) \right\rangle
= {\Phi_Q}\,\,\left( \sin {k_x}a
{\sigma^{1\,\,\alpha}_\beta}
+ i \sin {k_y}a
{\sigma^{2\,\,\alpha}_\beta}\right)
\end{equation}
As in the superconducting case,
more complicated order parameters are
possible, with all components of $\vec{\Delta}$
taking non-vanishing values.

For commensurate ordering, we can
follow the same logic as in (\ref{eqn:comm_con}).
The phases of the components of
${\vec{\Phi}_Q}(p)$ are constrained
in the same way as the singlet order parameters, as illustrated by the
$i$ in front of the second term in
(\ref{eqn:p-h_un_order}).

The orders discussed here can be generalized to
other $2D$ lattices and to $3D$ lattices. The
orbital wavefunctions $\sin {k_x}a$, etc.
will be replaced by representations of the 
point groups of these other lattices.

To each $T$-preserving singlet ordering, we can associate an
ordering of a pure spin model, in the same way that
spin-Peierls ordering is related to Peierls ordering:
\begin{equation}
\left\langle {\psi^{\alpha\dagger}}(k+Q,t)\,
{\psi_\alpha}(k,t) \right\rangle
\rightarrow 
\left\langle \vec{S}(k+Q,t)\,\cdot\,
\vec{S}(k,t) \right\rangle
\end{equation}
These spin orderings are states in which the
exchange energies are large along preferred directions.
These preferred diractions oscillate
from one lattice point to the next
with spatial frequency $\vec{Q}$.
The simplest case is spin-Peierls ordering,
in which the spins form dimers.
Another example is $d_{xy}$ ordering
of a spin model, which takes the form:
\begin{eqnarray*}
\left\langle \vec{S}(\vec{x},t)\,
\vec{S}(x+a{\hat x}+a{\hat y},t)\right\rangle +
\left\langle\vec{S}(\vec{x},t)\,
\vec{S}(x-a{\hat x}-a{\hat y},t)\right\rangle\cr -
\left\langle \vec{S}(\vec{x},t)\,
\vec{S}(x-a{\hat x}+a{\hat y},t)\right\rangle - 
\left\langle \vec{S}(\vec{x},t)\,
\vec{S}(x+a{\hat x}-a{\hat y},t)
\right\rangle \cr = -\frac{1}{4} 
\left({\Phi_Q}\,{e^{i\vec{Q}\cdot\vec{x}}}
+{\Phi_{-Q}}\,{e^{-i\vec{Q}\cdot\vec{x}}}\right)
\end{eqnarray*}
in analogy with (\ref{eqn:d_xy_def}).

\section{Model Hamiltonians}

We are primarily concerned in this paper with the
universal properties of the states introduced above.
We will not attempt to show that particular realistic
models of interacting electrons have
$p$- or $d$-wave density-wave ground states.
Rather, we will content ourselves with
discussing the types of interactions which favor
such orders and showing that they lead to
energetically favorable trial variational
wavefunctions for some idealized Hamiltonians.

The analog of the BCS reduced Hamiltonian for
singlet density-wave order is:
\begin{eqnarray}
\label{eqn:H_red}
H &=& \int \frac{{d^2}k}{(2\pi)^2}\,
\epsilon(k){\psi^{\alpha\dagger}}(k)
\,{\psi_{\alpha}}(k)\,\,-\cr & &
g\int \frac{{d^2}k}{(2\pi)^2}\,\frac{{d^2}k'}{(2\pi)^2}\,
\biggl[f(k)\,f(k')\,\times\cr
& &{\hskip 1 cm}\,{\psi^{\alpha\dagger}}(k+Q)
\,{\psi_{\alpha}}(k)\,
{\psi^{\beta\dagger}}(k')\,{\psi_{\beta}}(k'+Q)
\biggr]
\end{eqnarray}
In the triplet case, we replace the
four-fermion operator of (\ref{eqn:H_red}) by
\begin{eqnarray}
{\psi^{\alpha\dagger}}(k+Q)\,{\sigma^{a\,\,\beta}_{\,\,\alpha}}
\,{\psi_{\beta}}(k)\,
{\psi^{\gamma\dagger}}(k')\,{\sigma^{a\,\,\delta}_{\,\,\gamma}}
\,{\psi_{\delta}}(k'+Q)
\end{eqnarray}
We now introduce the variational wavefunction
\begin{eqnarray}
\label{eqn:trial_psi}
\Big|\Psi\Big\rangle = 
{\prod_{k,\alpha}}\left(
{u_{k,\alpha}}{\psi^{\alpha\dagger}}(k)
+ {v_{k,\alpha}}{\psi^{\alpha\dagger}}(k+Q)
\right)
\Big|0 \Big\rangle
\end{eqnarray}
Its energy can be minimized
if we take
\begin{equation}
{{\overline u}_{k,\alpha}}{v_{k,\alpha}} = 
\frac{g{\Phi_Q}\, f(k)}{\sqrt{{\left(\epsilon(k)
-\epsilon(k+Q)\right)^2}+
4{g^2}{\left|\Phi_Q\right|^2}{\left(f(k)\right)^2}}}
\end{equation}
in the singlet case and
\begin{equation}
{{\overline u}_{k,\alpha}}{\vec{\sigma}^{\,\,\beta}_{\alpha}}
{v_{k,\beta}} = 
\frac{g{\vec{\Phi}_Q}\, f(k)}{\sqrt{{\left(\epsilon(k)
-\epsilon(k+Q)\right)^2}+
4{g^2}{\left|\Phi_Q\right|^2}{\left(f(k)\right)^2}}}
\end{equation}
in the triplet case, and require $\Phi_Q$
to satisfy the gap equation:
\begin{equation}
g\int \frac{{d^2}k}{(2\pi)^2}\,
\frac{{\left(f(k)\right)^2}}{\sqrt{{\left(\epsilon(k)
-\epsilon(k+Q)\right)^2}+
4{g^2}{\left|\Phi_Q\right|^2}{\left(f(k)\right)^2}}}
= 1
\end{equation}
The reduced Hamiltonian has long-ranged interactions,
so the variational wavefunction is essentially correct.
We will now show that short-ranged Hamiltonians will
include terms of the form (\ref{eqn:H_red}),
and that the trial wavefunction (\ref{eqn:trial_psi})
is reasonable for these short-ranged Hamiltonians.

Consider, then, the following lattice model of interacting
electrons:
\begin{eqnarray}
H &=& -\,t {\sum_{<i,j>}}\left({c^\dagger_{i\sigma}}{c^{}_{j\sigma}}
+ {\rm h.c.}
\right)\, + \,U\, {\sum_i}{n_{i\uparrow}}{n_{i\downarrow}}\cr
& &\,+\, V{\sum_{<i,j>}}{n_{i}}{n_{j}}\cr
& &\,- \,{t_{c1}}{\sum_{<i,j>,<i',j>,i\neq i'}}{c^\dagger_{i\sigma}}
{c^{}_{j\sigma}}
{c^\dagger_{j\sigma}}{c^{}_{i'\sigma}}\cr
& &\,- \,{t_{c2}}{\sum_i}\Biggl[
\left({c^\dagger_{i+{\hat x},\sigma}}{c^{}_{i\sigma}}-
{c^\dagger_{i\sigma}}{c^{}_{i+{\hat x},\sigma}}\right)\times\cr
& &{\hskip 2 cm}
\left({c^\dagger_{i+{\hat x}+{\hat y},\sigma}}{c^{}_{i+{\hat y},\sigma}}-
{c^\dagger_{i+{\hat y},\sigma}}{c^{}_{i+{\hat x}+{\hat y},\sigma}}\right)\cr
& &{\hskip 2 cm} + x \rightarrow y \Biggr] 
\end{eqnarray}
The first two terms are the usual hopping, $t$, and on-site repulsion, $U$
of the Hubbard model. The third term is a nearest-neighbor repulsion, $V$.
The third and fourth terms lead to the correlated motion
of pairs of electrons. $t_{c1}$ hops an electron
from $i'$ to $j$ when $j$ is vacated by
an electron hopping to $i$. $t_{c2}$ hops nearest-neighbor pairs
in the same direction. Terms of this general form have been
discussed in \cite{Wheatley88,Chakravarty93,Hirsch89,Assaad97}
as a mechanism for superconductivity. As we will see below, they
not only favor superconductivity, but $p$- and $d$-wave
density wave order as well.

Fourier transforming the interaction terms into momentum space,
we see that terms of the form of the reduced interaction
(\ref{eqn:H_red}) are, indeed, present:
\begin{eqnarray}
{H_{\rm int}} &=& \int \frac{{d^2}k}{(2\pi)^2}\,\Biggl[
U\,{\psi^{\uparrow\dagger}}({k_1})\,{\psi_{\uparrow}}({k_2})
\,{\psi^{\downarrow\dagger}}({k_3})\,{\psi_{\downarrow}}({k_4})\cr
& & + \Biggl( 2V\,\Bigl(\cos\left({k^x_3}-{k^x_4}\right)a
+ \cos\left({k^y_3}-{k^y_4}\right)a\Bigr)\cr
& &\,\,\,\,\,\,\,-\,2{t_{c1}}\,\Bigl(\cos\left({k^x_1}-{k^x_4}\right)a
+ \cos\left({k^y_1}-{k^y_4}\right)a\cr
& &{\hskip 1.5 cm}+ 2\cos{k^x_1}a\cos{k^x_4}a
+ 2\cos{k^y_1}a\cos{k^y_4}a\Bigr)\cr
& &\,\,\,\,\,\,\,- \,2 {t_{c2}}\,\Bigl(\sin{k^x_1}a\sin{k^x_4}a
+\sin{k^y_1}a\sin{k^y_4}a\Bigr)
\Biggr)\cr
& &{\hskip 2 cm}\times\,\,{\psi^{\alpha\dagger}}({k_1})
\,{\psi_{\alpha}}({k_2})
\,{\psi^{\beta\dagger}}({k_3})\,{\psi_{\beta}}({k_4})
\Biggr]\cr
\end{eqnarray}

Let us now consider various candidate orderings
and the terms which favor or penalize them.
Antiferromagnetic order is favored by $U$
but penalized by $V$. Charge-density-wave
order is favored by $V$ but penalized by $U$.
$p$ and $d$-wave superconductivity are favored
by ${t_{c2}}$ and ${t_{c1}}$ respectively and penalized by $V$.
$p$ and $d$-density-wave order are favored by
${t_{c2}}$ and ${t_{c1}}$, respectively, and are both
favored by $V$. The density-wave states can be favored over
the others by taking $V$ large. The $p$- or $d$-wave
states can be favored by taking ${t_{c2}}$ or ${t_{c1}}$
large. To be more precise, the mean-field equations
for various ordered states read:
\begin{equation}
\lambda\int \frac{{d^2}k}{(2\pi)^2}\,
\frac{{\left(f(k)\right)^2}}{\sqrt{{\left(\epsilon(k)
-\epsilon(k+Q)\right)^2}+
4{\lambda^2}{\left|\Phi_Q\right|^2}{\left(f(k)\right)^2}}}
= 1
\end{equation}
where
\begin{eqnarray}
{\lambda_{dDW}} &=& 8V + 96{t_{c1}}\cr
{\lambda_{pDW}} &=& 4V + 16{t_{c1}} + 16{t_{c2}} \cr
{\lambda_{CDW}} &=& 16V + 96{t_{c1}} + 16{t_{c2}} - 2U\cr
{\lambda_{AF}} &=& 2U
\end{eqnarray}
Hence, the singlet $d_{{x^2}-{y^2}}$ density-wave
state will be the ground state if
\begin{eqnarray}
8{t_{c1}} < U-4V < 48{t_{c1}}\cr
8{t_{c2}} < 2V + 40{t_{c1}}\cr
\end{eqnarray}
while the singlet $p_x$ density-wave state
will be the ground state if
\begin{eqnarray}
6V + 40{t_{c1}} < U < 2V + 8{t_{c1}} + 8{t_{c2}}\cr
2V + 40{t_{c1}} < 8{t_{c2}} \cr
\end{eqnarray}
assuming that the van Hove singularities are
at the antinodes of the order parameters. Otherwise,
the $p$- and $d$-wave density wave states
will be favored over somewhat smaller regions
of parameter space.
By including spin-dependent interactions
such as $J\,{\vec{S}_i}\cdot{\vec{S}_i}$,
we can favor the
triplet $p$- or $d$-wave states.
Hence, it appears
that the orderings discussed in this paper
are viable. The detailed energetics at large
coupling strengths -- which surely
hold in physically interesting
systems -- are beyond the scope of this paper.

\section{Experimental Signatures}

We now turn to the question of the experimental
signatures of such states.
Since the order parameter changes sign
as the Fermi surface is circled, there is
no net CDW or SDW order which could be measured
in, for instance, neutron 
scattering\footnote{One can expect,
on general grounds, that incommensurate
singlet or triplet $p$- or $d$-wave
density-wave order at wavevector $\vec{Q}$
will induce CDW order at
$2\vec{Q}$ since a term of the
form ${\Phi_Q^2}\rho_{2Q}$ or
${\vec{\Phi}_Q}\cdot{\vec{\Phi}_Q}\rho_{2Q}$
is allowed by symmetry in the effective action.
Nevertheless, we may wish to distinguish such
a state from one which has only CDW order.}.
When another symmetry --
in addition to translational invariance --
is broken, this is easier.

Let's first consider broken time-reversal
symmetry. The commensurate singlet
${d_{{x^2}-{y^2}}}$ density wave state -- or
staggered flux state \cite{Kotliar88a,Marston89,Hsu91}
-- breaks $T$; there is an
alternating pattern of currents circulating about
each plaquette of the lattice. These currents produce an alternating
magnetic field measurable by $\mu$SR and, in principle,
by neutron scattering \cite{Hsu91}.
The magnitude of the current along a link of
the lattice will be:
\begin{equation}
j = \frac{e\,t}{\hbar}\,{\Phi_Q} \sim {10^{-5} {\rm Ampere}}\,\times\,{\Phi_Q}
\end{equation}
Now, ${\Phi_Q}$ is related to
the maximum of the gap according to
${\Delta_0}=g{\Phi_Q}$ where $g$ is the appropriate
coupling constant. Let us suppose, for the purposes
of illustration, that the formation of the ordered state
is driven by ${\lambda_{dDW}}$. Then, for ${\lambda_{dDW}}$ small,
${\Phi_Q} \sim (t/{\lambda_{dDW}}){e^{-({\rm const.})t/{\lambda_{dDW}}}}$.
Alternatively, we may take the high-$T_c$ context
as a guideline: observed gaps are $\sim 100-300K$,
while interactions such are $\sim 1 eV$. In this case,
we expect ${\Phi_Q}\sim {10^{-2}}$.
This translates to a magnetic field at the center
of each plaquette on the order of $10 G$.
The muons in a $\mu SR$ experiment might
see a lowwer field if they sit at points
of high symmetry or away from the plane.
The orbital magnetic moments are likely
to be dwarfed by local spin moments \cite{Hsu91}.
Incommensurate ordering may or may
not break $T$; if it does, the above analysis
applies.

In the $\Phi_Q^3\neq 0$, $\Phi_Q^1=\Phi_Q^2=0$ 
triplet ${d_{{x^2}-{y^2}}}$ density wave
state, there are counter-circulating currents
of up- and down-spin electrons. These currents cancel,
so there is no net current circulating
about each plaquette, but there is an alternating
pattern of spin currents circulating about each plaquette.
The checkerboard pattern of spin currents will
generate, via the spin-orbit coupling,
\begin{eqnarray}
{H_{SO}} &=& \int\frac{{d^2}k}{(2\pi)^2}\,\frac{{d^2}q}{(2\pi)^2}\,
\vec{E}(q)\cdot\left(2\vec{k}+\vec{q}\right)\times\cr
& &{\hskip 3 cm}{\psi^{\dagger\alpha}}(k+q)\,
{\vec{\sigma}_\alpha^{\,\,\,\beta}}{\psi_\beta}(k)
\end{eqnarray}
a quadrupolar electric field which is, in principle,
measurable in NQR experiments. With the above estimate
of the current, a nucleus with a non-zero
quadrupole moment would have an induced splitting of
order ${10 Hz}$.

We now turn to broken spin-rotational invariance,
characteristic of the triplet states. Since it
transforms non-trivially under the point
group of the square lattice, the
triplet order parameter ${\vec{\Phi}_Q}$ will
not couple to photons, neutrons, or nuclear spins
according to ${\vec{\Phi}_Q}\cdot\vec{F}$,
where $\vec{F}$ is, respectively,  $\vec{B}$, ${\vec{S}_N}$,
or $\vec{I}$. Said more physically,
the triplet ordered states do not have anomalous expectation
values for the spin density but, rather, for spin currents;
spin currents do not couple simply to these probes.
However, the order parameter ${\vec{\Phi}_Q}$ will couple to
such probes at second order since its square transforms
trivially under the point group.
Such a coupling will be of the form:
\begin{equation}
{H_{\rm probe}} = \int {d^2}x\,\left[
2\left(\vec{F}\cdot{\vec{\Phi}_Q}\right)
\left(\vec{F}\cdot{\vec{\Phi}^*_Q}\right)
-{\left|{\vec{\Phi}_Q}\right|^2}\,{F^2}\right]
\end{equation}
In the case of photons, this will lead
to $2$-magnon Raman scattering. The coupling to
nuclear spins couples directly to the nuclear quadrupole
moment, and will lead to a shift in the nuclear
quadrupole resonance frequencies.

In the presence of disorder, rotational
symmetry will be broken. Hence, there will be a
small coupling, proportional to the disorder strength,
of the Goldstone bosons to s-wave
probes such as NMR and neutron scattering.

\section{Gapless Fermionic Excitations}

The mean-field Hamiltonian is:
\begin{eqnarray}
H &=& {\int_{\rm B.Z.}} \frac{{d^2}k}{(2\pi)^2}\,\Biggl[
\epsilon(k)\,{\psi^{\alpha\dagger}}(k)\,
{\psi_\alpha}(k) +\cr & &{\hskip 2 cm}
g\,{\Phi_Q}f(k)\, {\psi^{\alpha\dagger}}(k+Q)\,
{\psi_\alpha}(k)
\Biggr]
\end{eqnarray}
If we define the four component object
$\chi_{A\alpha}$ according to
\begin{eqnarray}
\left(
\begin{array}{c}
{\chi_{1\alpha}}\\ {\chi_{2\alpha}}
\end{array}
\right)
= \left(
\begin{array}{c}
{\psi_\alpha}(k)\\
{\psi_\alpha}(k+Q)
\end{array}
\right)
\end{eqnarray}
then the mean-field Hamiltonian can be written
in the form:
\begin{eqnarray}
H &=& {\int_{\rm R.B.Z.}} \frac{{d^2}k}{(2\pi)^2}\,\,
{\chi^{\alpha\dagger}}(k)\Biggl(
\frac{1}{2}\left(\epsilon(k)-\epsilon(k+Q)\right){\tau_z}\,\, +\cr
& & {\hskip 1 cm } \Delta(k)\,{\tau_x}+
\frac{1}{2}\left(\epsilon(k)+\epsilon(k+Q)\right)
 \Biggr){\chi_{\alpha}}
\end{eqnarray}
The integral is over the reduced Brillouin zone. 
The $\tau$'s are Pauli matrices; the `flavor' index $A=1,2$ on which
they act has been suppressed. ${\Delta}(k)$ is defined by.
\begin{equation}
{\Delta}(k) \equiv {\Delta_0}f(k) \equiv g{\Phi_Q}f(k)
\end{equation}
The single-quasiparticle energies are:
\begin{eqnarray}
{E_\pm}(k) &=& \frac{1}{2}\left(\epsilon(k)+\epsilon(k+Q)\right)\,\pm\cr
& &\,\,\,\, \frac{1}{2}\sqrt{{\left(\epsilon(k)-\epsilon(k+Q)\right)^2}+
4{\Delta^2}(k)}
\end{eqnarray}

Let's consider the situation in which there is a node,
i.e. when the argument of the square root
vanishes (we discuss below the conditions under
which this occurs). For simplicity, we will 
consider the commensurate $\vec{Q}=(\pi/a,\pi/a)$
singlet $p_x$ density-wave state in a model with anisotropic
nearest-neighbor hopping:  
\begin{equation}
\label{disp}
\epsilon(k) = -2t\left( r \cos{k_x}a + \cos{k_y}a \right)
\end{equation}
with $r<1$. The mean-field quasiparticle energies are:
\begin{equation}
E(k) = \pm \sqrt{{4t^2}{\left( r \cos{k_x}a + \cos{k_y}a \right)^2}
+ {\Delta_0}^2{\sin^2}{k_x}a}
\end{equation}
There is a node at ${k_x}=0$, ${k_y}a = \arccos(-r)$. Expanding about
this node,
\begin{equation}
E(q) = \pm \sqrt{{v_x^2}{q_x^2} + {v_y^2}{q_y^2}}
\end{equation}
with momenta $\vec{q}$ now measured from the node
and
\begin{equation}
{v_x} = {\Delta_0}a\, , {\hskip 0.7 cm}
{v_y} = 2ta\sqrt{1-{r^2}}
\end{equation}
The effective Lagrangian for the quasiparticles near
the nodes can be written:
\begin{equation}
{{\cal L}_{\rm eff}} = {\chi^{\alpha\dagger}}
\left( {\partial_\tau} - {\tau_z}{v_y}i{\partial_y}
- {\tau_x}{v_x}i{\partial_x}\right)
{\chi_{\alpha}}
\end{equation}
Terms which break the nesting of the
Fermi surface, such as the chemical potential
or next-neighbor hopping, open hole pockets at the nodes:
\begin{equation}
{{\cal L}_\mu} = - \mu\, {\chi^{\alpha\dagger}}{\chi_{\alpha}}
\end{equation}

We now turn to the question of when
a $p$- or $d$-wave density wave
will have nodal excitations.
Let's again begin with the commensurate
$\vec{Q}=(\pi/a,\pi/a)$ singlet $p_x$ density-wave
state:
\begin{equation}
\left\langle {\psi^{\alpha\dagger}}(k+Q,t)\,
{\psi_\beta}(k,t) \right\rangle
= i\left|{\Phi_Q}\right| \sin {k_x}a\,\,
{\delta^\alpha_\beta}
\end{equation}
in a system in which the Fermi surface
is nested at $\vec{Q}$.
This state will have gapless excitations
if the nodal line ${k_x}=0$ crosses the
Fermi surface. For an open Fermi surface, this need
not be the case. In an anisotropic nearest-neighbor
tight-binding model, Eq. (\ref{disp}), 
with $r>1$, the Fermi surface at half-filling is
an open Fermi surface which does not cross the
line ${k_x}=0$. Consequently, there are no
gapless excitations. For $r<1$, however, the Fermi surface
does cross the line ${k_x}=0$, and there are gapless
excitations.

Are there any thermodynamic singularities
at the transition at which gapless excitations
occur? To answer this question, let us consider
the mean-field ground state energy:
\begin{eqnarray}
{E_0} &=& {\int_{\rm R.B.Z.}} \frac{{d^2}k}{(2\pi)^2}\, E(k)\cr
&=& {\int_{\rm R.B.Z.}}\frac{{d^2}k}{(2\pi)^2}\,
\sqrt{{\epsilon^2}(k)+ {\Delta^2}(k)}
\end{eqnarray}
The first and second derivatives of
${E_0}$ are continuous.
However, the third derivative of the ground state
energy with respect to $r$ contains a term of the form:
\begin{eqnarray}
\frac{{\partial^3}{E_0}}{\partial r^3} =
{\int_{\rm R.B.Z.}}\frac{{d^2}k}{(2\pi)^2}\,
\frac{8\epsilon(k)\,{t^3}{\cos^3}{k_x}a}{\left({\epsilon^2}(k)+
{\Delta^2}(k)\right)^{3/2}} + \ldots
\end{eqnarray}
This term diverges if there is a node on the
Fermi surface, but is finite otherwise.
Hence, the phase with a node
on the Fermi surface is a {\it critical line}
with a singular third derivative of the ground state
energy. We will call this phase the
`critical phase' of the $p_x$ density wave.
Note that the second derivative of the ground
state energy is everywhere continuous but
nowhere differentiable in the critical
phase. It is separated by a third-order phase transition
from the phase with no gapless excitations, the non-critical
phase of the $p_x$ density wave.

How does this observation generalize (a)
away from half-filling and to non-nested Fermi surfaces;
and (b) to $d$-wave
and/or incommensurate ordering?
To answer (a), let's change the chemical potential
in order to move away from half-filling. Now,
\begin{eqnarray}
\frac{{\partial^3}{E_0}}{\partial r^3} =
{\int_{E(k)<\mu}}\frac{{d^2}k}{(2\pi)^2}\,
\frac{8\epsilon(k)\,{t^3}{\cos^3}{k_x}a}{\left({\epsilon^2}(k)+
{\Delta^2}(k)\right)^{3/2}} + \ldots
\end{eqnarray}
Below half-filling, the denominator never diverges.
Hence, the system is always in the non-critical phase,
despite the fact that there are gapless excitations.
As the chemical potential is increased, the system crosses
a third-order phase transition and enters the
critical phase. Above half-filling, it is always
in the critical phase.

Suppose, now, that we allow
next-nearest neighbor hopping $t'$, thereby
spoiling nesting. The ground state energy
is given by
\begin{eqnarray}
{E_0} &=& {\int_{E(k)<\mu}} \frac{{d^2}k}{(2\pi)^2}\, E(k)\cr
&=& {\int_{E(k)<\mu}}\frac{{d^2}k}{(2\pi)^2}\,
\Biggl[\frac{1}{2}\left(\epsilon(k)+\epsilon(k+Q)\right)-\cr
& & {\hskip 2 cm}
\frac{1}{2}\sqrt{{\left(\epsilon(k)-\epsilon(k+Q)\right)^2}+
4{\Delta^2}(k)}\biggr]
\end{eqnarray}
and
\begin{eqnarray*}
\frac{{\partial^3}{E_0}}{\partial r^3} &=&
{\int_{E(k)<\mu}}\frac{{d^2}k}{(2\pi)^2}\,
\frac{4\epsilon(k)\,{t^3}{\cos^3}{k_x}a}{\left(
{\left(\epsilon(k)-\epsilon(k+Q)\right)^2}+
4{\Delta^2}(k)\right)^{3/2}}\cr & & \,\,\,+ \ldots
\end{eqnarray*}
This diverges if the nodal line of ${\Delta}(k)$ crosses
the curve $\epsilon(k)=\epsilon(k+Q)$
and this crossing point lies below the
chemical potential. In such a case, the
system is in the critical phase.
Regardless of the details of the band structure, the
curve $\epsilon(k)=\epsilon(k+Q)$ is determined
by symmetry for commensurate $\vec{Q}$:
it is the set of points for
which $\vec{k}$ and $\vec{k}+\vec{Q}$ are
related by a symmetry of the square lattice.
For $\vec{Q}=(\pi/a,0)$, $\epsilon(k)=\epsilon(k+Q)$
if ${k_x}=\pm\pi/2a$. For $\vec{Q}=(\pi/a,\pi/a)$,
$\epsilon(k)=\epsilon(k+Q)$ if ${k_x}\pm{k_y}=\pm\pi/a$.
A $d$-wave density-wave
will always have nodal lines which cross
the curve $\epsilon(k)=\epsilon(k+Q)$; a $p$-wave density wave may
or may not.
If there is no crossing point, or the crossing point is not
below the chemical potential, the system is in the
non-critical phase. Again, the non-critical phase
can have gapless excitations.

In the case of incommensurate ordering, similar
considerations hold. Let us suppose that the Fermi
surface is nested at incommensurate $\vec{Q}$,
i.e. if $\vec{k}$ is on the Fermi surface,
then $\vec{k}+\vec{Q}$ is as well, and
$\epsilon(k)=\epsilon(k+Q)=\mu$.
If the Fermi surface intersects the nodal
lines of $\Delta(k)$, then there will be
gapless nodal excitations. If the chemical potential
is now lowered or the hopping parameters are
changed, so that the Fermi surface is
no longer perfectly nested, then the nodes will open
into hole pockets. Again, as the nesting
condition is approached, a third-order
phase transition will occur, as in the commensurate case.

In summary, the system will be in a `critical'
state if nodal points are at or below the Fermi surface.
Otherwise, the system will be `non-critical', whether or
not there other gapless excitations. The transition
between these two states and the entire critical
phase is characterized, in mean-field-theory, by
a diverent third derivative of the ground state
energy. There is no reason to mistrust mean-field
theory since there aren't strong order parameter
fluctuations which might destabilize out
calculations.

\section{Transitions to Superconducting
States}

As Zhang \cite{Zhang97} has recently emphasized, enhanced symmetry
can be dynamically generated at a critical point
between two different ordered electronic
states. The focus of that work was a critical point
between an antiferromagnet and a $d_{{x^2}-{y^2}}$
superconductor. In earlier work, Yang identified an
$SU(2)$ symmetry (which, together with $SU(2)$
spin-rotational symmetry, trivially forms
an $O(4)=SU(2)\times SU(2)\times {Z_2}$) which is an exact
symmetry of the Hubbard model at half-filling
with $\mu=U/2$. This symmetry would be dynamically generated
at a critical point between a CDW and an $s$-wave
superconductor. We now consider the modification of
this idea to $p$- and $d$-wave ordering.

We first consider a transition at
half-filling between a singlet
commensurate $d_{{x^2}-{y^2}}$ density-wave
and a $d_{{x^2}-{y^2}}$ superconductor. We group the two order
parameters into a vector\footnote{We will use underlined lowercase
Roman letters such as
${\underline i}=1,2,3$ to denote pseudospin triplet indices
and uppercase Roman letters to denote peudospin doublet
indices $A=1,2$. Lowercase Roman indices $a=1,2,3$
will be vector indices (i.e. real spin triplet
indices) and Greek letters $\alpha=1,2$ will
be used for real spin $SU(2)$ spinor indices.
Pauli matrices $\tau^{\underline i}$ will be used
for pseudospin, while $\sigma^a$ will be reserved for
spin.}:
\begin{eqnarray}
{\Phi_{\underline i}}(q)\,f(k) = \left(
\begin{array}{c}
\sqrt{2}\,{\rm Re}\left\{\left\langle
{\psi_\uparrow^\dagger}(k+\frac{q}{2})\,
{\psi_\downarrow^\dagger}(-k+\frac{q}{2}) \right\rangle\right\}\\
\sqrt{2}\,{\rm Im}\left\{\left\langle
{\psi_\uparrow^\dagger}(k+\frac{q}{2})\,
{\psi_\downarrow^\dagger}(-k+\frac{q}{2}) \right\rangle\right\}\\
i\left\langle {\psi^{\alpha\dagger}}(k+Q+\frac{q}{2})\,
{\psi_\alpha}(k-\frac{q}{2}) \right\rangle
\end{array}
\right)
\end{eqnarray}
If, following Yang \cite{Yang89}, we introduce the
following $SU(2)$ generators which we will
call pseudospin $SU(2)$
\begin{eqnarray}
{O^3} &=& {\int_{\rm R.B.Z.}}\frac{{d^2}k}{(2\pi)^2}\,\,
\biggl({\psi^{\alpha\dagger}}(k)\,
{\psi_\alpha}(k)\,\, +\cr
& & {\hskip 2.5 cm} {\psi^{\alpha\dagger}}(k+Q)\,
{\psi_\alpha}(k+Q)\biggr)\cr
{O^+} &=& {\int_{\rm R.B.Z.}}\frac{{d^2}k}{(2\pi)^2}\,\,
i{\psi_\uparrow^\dagger}(k)\,
{\psi_\downarrow^\dagger}(-k+Q)\cr
{O^-} &=& {\int_{\rm R.B.Z.}}\frac{{d^2}k}{(2\pi)^2}\,\,
i{\psi_\uparrow}(k)\,
{\psi_\downarrow}(-k+Q)
\end{eqnarray}
then the order parameters form a triplet
under this $SU(2)$,
\begin{eqnarray}
\left(
\begin{array}{c}
{\Phi_+}(q)\,f(k)\\ {\Phi_0}(q)\,f(k) \\ {\Phi_-}(q)\,f(k)
\end{array}
\right)
= \left(
\begin{array}{c}
-\left\langle {\psi_\uparrow^\dagger}(k+\frac{q}{2})\,
{\psi_\downarrow^\dagger}(-k+\frac{q}{2}) \right\rangle\\
i\left\langle {\psi^{\alpha\dagger}}(k+Q+\frac{q}{2})\,
{\psi_\alpha}(k-\frac{q}{2}) \right\rangle\\
\left\langle {\psi_\uparrow}(k+\frac{q}{2})\,
{\psi_\downarrow}(-k+\frac{q}{2}) \right\rangle
\end{array}
\right)
\end{eqnarray}
There is a small but important difference
between our pseudospin $SU(2)$ and
Yang's \cite{Yang89}: the factors of
$i$ in the definitions of ${O^\pm}$.
These are necessary since a commensurate
$d_{{x^2}-{y^2}}$ density-wave
breaks $T$, while a $d_{{x^2}-{y^2}}$ superconductor
does not. Consequently, our pseudospin $SU(2)$
does not commute with $T$, which
is an inversion followed by a rotation by
$\pi$ about the ${\underline 3}$-axis.

The electron fields transform as
a doublet under the pseudospin $SU(2)$
as well as the spin $SU(2)$. We will
group them into $4$-component objects
$\Psi_{A\alpha}$, where $A$ is the pseudospin
index, $A=1,2$, and $\alpha$ is the spin index,
$\alpha=\uparrow,\downarrow$:
\begin{eqnarray}
\left(
\begin{array}{c}
{\Psi_{1\alpha}}\\ {\Psi_{2\alpha}}
\end{array}
\right)
= \left(
\begin{array}{c}
{\psi_\alpha}(k)\\
i{\epsilon_{\alpha\beta}}{\psi^{\beta\dagger}}(-k+Q)
\end{array}
\right)
\end{eqnarray}

A `microscopic' Hamiltonian which
is $O(4)$ invariant can be written down:
\begin{equation}
H = {H_0} + {H_{int}}
\end{equation}
\begin{eqnarray}
{H_0} &=& {\int_{\rm R.B.Z.}}\frac{{d^2}k}{(2\pi)^2}\,
\epsilon(k)\,{\Psi^{A\alpha\dagger}}\,{\Psi_{A\alpha}}
\end{eqnarray}
if ${H_{int}}$ is given by\footnote{We have only written
down the quartic terms; higher-order $O(4)$ invariants also
exist, but they are irrelevant at weak coupling.}:
\begin{eqnarray*}
{H_{int}} &=& \int\frac{{d^2}q}{(2\pi)^2}\,
\biggl[{u^{(0,0)}}\,{\lambda^{(0,0)}}(q)\,{\lambda^{(0,0)}}(q)+
\cr
& & {u^{(1,0)}}\,{\lambda_{\underline i}^{(1,0)}}(q)
\,{\lambda_{\underline i}^{(1,0)}}(q)+
{u^{(1,0)}}\,{\lambda_{{\underline i}\,Q}^{(1,0)}}(q)\,
{\lambda_{{\underline i}\,Q}^{(1,0)}}(q)+\cr
& & {u^{(0,1)}}\,{\lambda_a^{(0,1)}}(q)\, {\lambda_a^{(0,1)}}(q) +
{u_Q^{(0,1)}}\,{\lambda_{a\,Q}^{(0,1)}}(q)\,
{\lambda_{a\,Q}^{(0,1)}}(q)+\cr
& & {u^{(1,1)}}\,{\lambda_{{\underline i}a}^{(1,1)}}(q)
\,{\lambda_{{\underline i}a}^{(1,1)}}(q)+
{u_Q^{(1,1)}}\,{\lambda_{{\underline i} a\,Q}^{(1,1)}}(q)\,
{\lambda_{{\underline i} a\,Q}^{(1,1)}}(q)\biggr]
\end{eqnarray*}
where
\begin{eqnarray}
{\lambda^{(0,0)}} &=& \int\frac{{d^2}k}{(2\pi)^2}\,f(k)\,
{\Psi^{A\alpha\dagger}}\left(k+\frac{q}{2}\right)
\,{\Psi_{A\alpha}}\left(k-\frac{q}{2}\right)\cr
{\lambda_{\underline i}^{(1,0)}} &=& \int\frac{{d^2}k}{(2\pi)^2}\,
f(k)\,{\Psi^{A\alpha\dagger}}
\left(k+\frac{q}{2}\right)\,\times\cr
& &{\hskip 3.5 cm}{\tau^{\,B}_{{\underline i}\,A}}
{\Psi_{B\alpha}}\left(k-\frac{q}{2}\right)\cr
{\lambda_{{\underline i}\,Q}^{(1,0)}} &=& \int\frac{{d^2}k}{(2\pi)^2}\,
f(k)\,{\epsilon^{\alpha\beta}}
{\Psi_{C\alpha}}\left(k+\frac{q}{2}\right){\epsilon^{CA}}\,\times\cr
& &{\hskip 3.5 cm}
{\tau^{\,B}_{{\underline i}\,A}}
{\Psi_{B\beta}}\left(-k+\frac{q}{2}\right)\cr
{\lambda_a^{(0,1)}}&=& \int\frac{{d^2}k}{(2\pi)^2}\,f(k)\,
{\Psi^{A\alpha\dagger}}\left(k+\frac{q}{2}\right)
{\sigma^{\,\beta}_{{\underline i}\,\alpha}}\,\times\cr
& &{\hskip 3.5 cm}
{\Psi_{A\beta}}\left(k-\frac{q}{2}\right)\cr
{\lambda_{a\,Q}^{(0,1)}} &=& \int\frac{{d^2}k}{(2\pi)^2}\,
f(k)\,{\epsilon^{AB}}
{\Psi_{A\gamma}}\left(k+\frac{q}{2}\right){\epsilon^{\gamma\beta}}
\,\times\cr & &{\hskip 3.5 cm}
{\sigma^{\,\beta}_{{\underline i}\,\alpha}}
{\Psi_{B\beta}}\left(-k+\frac{q}{2}\right)\cr
{\lambda_{{\underline i}a}^{(1,1)}}&=& \int\frac{{d^2}k}{(2\pi)^2}\,
f(k)\,{\Psi^{A\alpha\dagger}}\left(k+\frac{q}{2}\right)
{\tau^{\,B}_{{\underline i}\,A}}\,\times\cr
& &{\hskip 3.5 cm}
{\sigma^{\,\beta}_{{\underline i}\,\alpha}}
{\Psi_{B\beta}}\left(k-\frac{q}{2}\right)\cr
{\lambda_{{\underline i}a\,Q}^{(1,1)}} &=& \int\frac{{d^2}k}{(2\pi)^2}\,
f(k)\,{\Psi_{C\gamma}}\left(k+\frac{q}{2}\right){\epsilon^{CA}}
{\tau^{\,B}_{{\underline i}\,A}}\,\times\cr
& &{\hskip 3.5 cm}{\epsilon^{\gamma\beta}}
{\sigma^{\,\beta}_{{\underline i}\,\alpha}}
{\Psi_{B\beta}}\left(-k+\frac{q}{2}\right)
\end{eqnarray}

These `microscopic' Hamiltonians describe electrons
at half-filling with a nested Fermi surface
and interactions which favor
density-wave and superconducting order equally.
In other words, they describe a critical point
at half-filling between a $d_{{x^2}-{y^2}}$ density-wave
and a $d_{{x^2}-{y^2}}$ superconductor.
Near the critical point, we
can focus on the low-energy degrees of freedom:
the Goldstone modes and the nodal
fermionic excitations.
We can write down an $O(4)$ invariant action
for this:
\begin{eqnarray}
\label{eqn:o(4)_Lag}
{S_{\rm eff}} &=&
%  {\hskip 7 cm}\cr & &
 \int d\tau\,\frac{{d^2}k}{(2\pi)^2}\,
{\Psi^{A\alpha^\dagger}}(k) \left({\partial_\tau}
 -  \epsilon(k)\right) {\Psi_{A\alpha}}(k)\,+\cr
& &{\hskip -0.4 cm}i\,g \int d\tau\,\frac{{d^2}k}{(2\pi)^2}\,\frac{{d^2}q}{(2\pi)^2}
\,\,{\Phi_{\underline i}}(q)\,f(k)\,\times\cr
& &{\hskip 0.6 cm}\Bigl[
{\epsilon^{\alpha\beta}}{\Psi_{C\alpha}}\left(k+\frac{q}{2}\right)
{\epsilon^{CA}}
{\tau^{{\underline i}\,B}_A} {\Psi_{B\beta}}\left(-k+\frac{q}{2}\right)\,+\cr
& &{\hskip 0.8 cm}{\epsilon_{\alpha\beta}}
{\Psi^{A\alpha\dagger}}\left(k+\frac{q}{2}\right)
{\tau^{{\underline i}\,B}_A} {\epsilon^{BC}}
{\Psi^{B\beta\dagger}}\left(-k+\frac{q}{2}\right)\Bigl] \cr
& &+\,\int d\tau {d^2}x\,\left(
{\left({\partial_\mu}{\Phi_{\underline i}}\right)^2}+
\frac{1}{2}\,r\,{\Phi_{\underline i}}{\Phi_{\underline i}}
+ \frac{1}{4!}\,u\,
{\left({\Phi_{\underline i}}{\Phi_{\underline i}}\right)^2}
\right)
\end{eqnarray}
In this Lagrangian, we have rescaled all of the velocities
and stiffnesses to $1$. In general, these
quantities will be different -- breaking the $O(4)$
symmetry -- and this cannot be done. Symmety-breaking
terms will be briefly addressed below.

The transition between the $d_{{x^2}-{y^2}}$ density-wave
and the $d_{{x^2}-{y^2}}$ superconductor
is driven by a pseudospin-$2$ symmetry-breaking field,
which we will call $u$.
\begin{eqnarray}
{{\cal L}_{u}} &=& u\,
\left( {\Phi_0^2} + {\Phi_+}{\Phi_-}\right)\cr
&=& u\,
\left( {\Phi_{\underline 3}^2} - {\Phi_{\underline 1}^2}
- {\Phi_{\underline 2}^2}\right)
\end{eqnarray}
For $u<0$, the $3$-axis is an easy axis
and the $d_{{x^2}-{y^2}}$ density-wave state is
favored; for $u>0$, the $1-2$-plane is an easy plane
and the $d_{{x^2}-{y^2}}$ superconducting state is
favored.

We can move away from a nested Fermi surface
by tuning the chemical potential or a next-neighbor
hopping parameter. Such effects are encapsulated by a
pseudospin-$1$ symmetry-breaking term:

\begin{eqnarray}
{S_{\mu}} &=& \,\mu\, {O^3}\cr
&=&\int d\tau\,{d^2}x\,\left(
{\epsilon_{{\underline i}{\underline j}}}
{\Phi_{\underline i}}{\partial_\tau} {\Phi_{\underline j}}
+ {\Psi^\dagger} {\tau^{\underline 3}} \Psi \right)
\end{eqnarray}

where $O^3$ is the pseudospin $SU(2)$
generator defined above.
If $u=0$, $\mu$ will immediately force
the pseudospin into the $1-2$ plane -- i.e.
the supercondcutor will be favored.
If $u<0$, the $d_{{x^2}-{y^2}}$ density-wave state will
be favored until ${\mu_c} \propto (\sqrt{-u})$. At this
point, a first-order phase transition --
the pseudospin-flop transition -- will occur
at which the pseudospin switches from an easy-axis
phase to an easy-plane phase.
If we allow ${\Phi_0}$ to have a different
velocity than ${\Phi_\pm}$, then this first order phase
transition can become two second order phase transitions.
Depending on the values of these parameters and the strength of quantum
fluctuations, the intervening phase can either have both types of order
or neither.

The critical point occurs when the jump in
${\Phi}$ is tuned to zero. Hence, it is a tricritical
point. At such a critical point, $O(4)$-breaking
terms can scale to zero. The critical point
and the quantum critical region
\cite{Chakravarty89,Sachdev96} are described by the physics
of the critical fluctuations coupled
to nodal fermionic excitations. By arguments similar to
those of \cite{Balents98}, the nodal fermions
are neutral, spin-$1/2$ objects.
A more detailed analysis will
be given elsewhere \cite{Nayak00b}.

Similar conclusions can be drawn for
$d_{xy}$ and ${d_{{x^2}-{y^2}}}+id_{xy}$ transitions;
the latter case is particularly simple
since there are no fermions.
In the case of transitions between
$p_x$-wave density-wave and superconducting states,
the order parameters are both pseudospin
{\it and} spin-triplets. Hence, the effective
field theory for such a transition takes the form:
\begin{eqnarray*}
{S_{\rm eff}} &=& {\hskip 7 cm}\cr
& & \int d\tau\,\frac{{d^2}k}{(2\pi)^2}\,
{\Psi^{A\alpha^\dagger}} \left({\partial_\tau}
 -  \epsilon(k)\right) {\Psi_{A\alpha}}\,+\cr
& &{\hskip -1 cm}
i\,g \int d\tau\,\frac{{d^2}k}{(2\pi)^2}\,\frac{{d^2}q}{(2\pi)^2}
\,\,{\Phi^a_{\underline i}}(q)\,f(k)\,\times\cr
& & \Bigl[
{\epsilon^{\gamma\alpha}}{\sigma^{a\,\beta}_\alpha}
{\Psi_{C\gamma}}\left(k+\frac{q}{2}\right){\epsilon^{CA}}
{\tau^{a\,B}_A} {\Psi_{B\beta}}\left(-k+\frac{q}{2}\right)\,+\cr
& &{\hskip 0.2 cm}{\sigma^{a\,\beta}_\alpha}{\epsilon_{\beta\gamma}}
{\Psi^{A\alpha\dagger}}\left(-k+\frac{q}{2}\right)
{\tau^{a\,B}_A} {\epsilon^{BC}}
{\Psi^{B\gamma\dagger}}\left(-k+\frac{q}{2}\right)\Bigl] \cr
& &+\,\int d\tau {d^2}x\,\left(
{\left({\partial_\mu}{\Phi_{\underline i}}\right)^2}+
\frac{1}{2}\,r\,{\Phi^a_{\underline i}}{\Phi^a_{\underline i}}
+ \frac{1}{4!}\,u\,
{\left({\Phi^a_{\underline i}}{\Phi^a_{\underline i}}\right)^2}
\right)\cr
\end{eqnarray*}
where
\begin{eqnarray}
\left(
\begin{array}{c}
{\Phi_{\underline 1}^a}\\ {\Phi_{\underline 2}^a}
\\ {\Phi_{\underline 3}^a}
\end{array}
\right) = \left(
\begin{array}{c}
\sqrt{2}\,{\rm Re}\left\{\left\langle {\psi_\gamma^\dagger}(k)\,
{\epsilon^{\gamma\alpha}}{\sigma^{a\,\beta}_\alpha}\,
{\psi_\beta^\dagger}(-k) \right\rangle\right\}\\
\sqrt{2}\,{\rm Im}\left\{\left\langle {\psi_\gamma^\dagger}(k)\,
{\epsilon^{\gamma\alpha}}{\sigma^{a\,\beta}_\alpha}\,
{\psi_\beta^\dagger}(-k) \right\rangle\right\}\\
i\left\langle {\psi^{\alpha\dagger}}(k+Q)\,
{\sigma^{a\,\beta}_\alpha}\,
{\psi_\beta}(k) \right\rangle
\end{array}
\right)
\end{eqnarray}

\section{Discussion}

In this paper, we have discussed
the properties of ordered states
in which particle-hole pairs with
non-zero angular momentum condense.
These states generalize charge-
and spin-density wave states in the
same way that $p$- and $d$-wave superconductors
generalize $s$-wave superconductivity.
However, unlike in the superconducting case --
where the Meissner effect follows directly
from the broken symmetry, irrespective of
the pairing channel -- the angular variation
of the condensate makes $p$- and $d$-wave
density-wave ordering difficult to detect.
Experiments seeking to uncover such order
must (a) be sensitive to spatial variations
of kinetic energy or currents or (b) measure
higher-order correlations of the charge
or spin density. We explained how $\mu$SR, neutron
scattering, NQR, and Raman scattering can be used in this
regard. Impurities, which break rotational invariance,
would cause admixture of $p$- or $d$-wave
ordering with $s$-wave ordering. It is natural
to wonder whether experiments which appear to detect
SDW order should be re-examined
to see if they have actually uncovered
$p$- or $d$-wave order
which, as a result of impurities,
is masquerading as $s$-wave order.

As in the superconducting case,
the non-trivial pairing symmetry can lead
to the existence of nodal excitations.
As parameters such as the chemical potential
or next-neighbor hopping are varied, nodal excitations
appear at a transition which is third-order in
mean field theory. The `phase' with nodal excitations
is always critical.

The analogies between $p$- and $d$-wave
density-wave ordering and $p$- and $d$-wave
superconductivity begs the question:
what is the nature of a phase transition
between such states? In answering this question,
we are led to one of the motivations of this work.
The pseudo-gap regime of the cuprate superconductors
exhibits some properties which can be associated
with $d_{{x^2}-{y^2}}$ ordering.
One explanation is that some features
of the $d_{{x^2}-{y^2}}$
superconducting state have been inherited.
However, it is natural to inquire whether
the physics of this regime could also
be determined in part by
proximity to a $d_{{x^2}-{y^2}}$ density-wave
state or the transition between the
density-wave and superconducting states.
In other words, we ask whether the
physics of the pseudo-gap regime
should be described by a theory
which incorporates fluctuations
between $d_{{x^2}-{y^2}}$ density-wave
and superconducting states.
In this connection, we note that
the physics of the critical point between
$d_{{x^2}-{y^2}}$ density-wave and superconducting 
states bears a rough resemblance to that
of the $SU(2)$ mean field theory of the
$t-J$ model \cite{Wen96,Lee98}.
In that theory, the gauge field
parametrizes fluctuations between the $d_{{x^2}-{y^2}}$
density-wave and superconducting states, a role played
in our analysis by the Goldstone bosons of the
$O(4)$ effective theory. We also note
that the Nodal Liquid state \cite{Balents98,Balents99a,Balents99b}
shares many features of the 
$d_{{x^2}-{y^2}}$ density-wave
and superconducting states. These issues
and their possible relevance to the
cuprates will be further explored elsewhere.

I would like to thank Sudip Chakravarty,
Stuart Brown, Subir Sachdev,
and, especially, Dror Orgad
for discussions.

%\bibliography{corr}
%\bibliographystyle{prsty}

\end{document}